\documentclass{SciPost}

\usepackage{amssymb}
\usepackage{amsmath}
\usepackage{physics}
\usepackage{xspace}
\usepackage{color}
\usepackage{hyperref}
\usepackage{mathtools}
\usepackage{comment}
\usepackage[normalem]{ulem}

\newcommand{\HH}{{\cal H}}
\newcommand{\VV}{{\cal V}}
\newcommand{\e}{{\mathrm{e}}}
\newcommand{\im}{{\mathrm{i}}}
\newcommand{\kF}{{k_\mathrm{F}}}
\newcommand{\Ssp}{{S_\mathrm{sp}}}

\newcommand{\Eqref}[1]{Eq.~\eqref{#1}}

\DeclareMathOperator{\Span}{span}

\hypersetup{
    colorlinks,
    linkcolor={red!50!black},
    citecolor={blue!50!black},
    urlcolor={blue!80!black}
}

\usepackage[bitstream-charter]{mathdesign}
\DeclareSymbolFont{usualmathcal}{OMS}{cmsy}{m}{n}
\DeclareSymbolFontAlphabet{\mathcal}{usualmathcal}

\fancypagestyle{SPstyle}{
\fancyhf{}
\lhead{\colorbox{scipostblue}{\bf \color{white} ~SciPost Physics Core }}
\rhead{{\bf \color{scipostdeepblue} ~Submission }}

\fancyfoot[C]{\textbf{\thepage}}
}

\begin{document}

\pagestyle{SPstyle}

\begin{center}{\Large \textbf{\color{scipostdeepblue}{
Interaction-driven dynamics in graphene flakes as a benchmark for quantum simulation
}}}\end{center}

\begin{center}\textbf{
Fabian Eickhoff\textsuperscript{1$\star$},
Satoshi Ejima\textsuperscript{2},
Lukas Windgätter\textsuperscript{2},
Florian G. Eich\textsuperscript{3},
Hannah Rittich\textsuperscript{3},
Sebastian Zanker\textsuperscript{3} and
Peter Schmitteckert\textsuperscript{3$\dagger$}
}\end{center}

\begin{center}
{\bf 1} Institute of Software Technology, German Aerospace Center,  51147 Cologne, Germany
\\
{\bf 2} Institute of Software Technology, German Aerospace Center, 22529 Hamburg, Germany
\\
{\bf 3} HQS Quantum Simulations GmbH, 76131 Karlsruhe, Germany
\\[\baselineskip]
$\star$ \href{fabian.eickhoff@dlr.de}{\small fabian.eickhoff@dlr.de}\,,\quad
$\dagger$ \href{Peter.Schmitteckert@quantumsimulations.de}{\small Peter.Schmitteckert@quantumsimulations.de}
\end{center}

\section*{\color{scipostdeepblue}{Abstract}}
\textbf{\boldmath{%
We study interaction-driven ultrafast dynamics in finite graphene flakes following an optical pump quench in an interacting tight-binding model.
By comparing exact real-time evolution with simulations restricted to particle–hole excitation subspaces, we assess when relaxation can be captured by low-order many-body processes and when this is not sufficient. The single-particle orbital entropy provides a compact diagnostic for dynamic correlation growth.
For the systems studied here, periodic graphene flakes are well described by low-order excitations, whereas confined geometries require
substantial higher-order contributions even for relatively small interaction strengths. The quench protocol combines simple initial-state preparation with strongly correlated dynamics, identifying a promising benchmark problem for future quantum-computing simulations.
}}

\vspace{\baselineskip}



\vspace{10pt}
\noindent\rule{\textwidth}{1pt}
\tableofcontents
\noindent\rule{\textwidth}{1pt}
\vspace{10pt}


\section{Introduction}
Ultrafast carrier relaxation in graphene has attracted sustained interest because it combines fundamental many-body physics with direct relevance for optoelectronic and quantum-device applications \cite{CastroNeto2009,DasSarma2011,Basov2014}.
Owing to its linear low-energy dispersion and reduced dimensionality, graphene supports an electron fluid in which Coulomb interaction rapidly redistribute energy and momentum after optical or electrical excitation  \cite{CastroNeto2009,Mueller2009}.
A large body of pump--probe, THz, and time- and angle-resolved photoemission experiments has established that non-thermal carrier populations relax on extremely short timescales, typically of the order of a few tens of femtoseconds, before slower phonon-assisted cooling takes over \cite{Dawlaty2008,Breusing2011,Brida2013,Gierz2013,Gierz2015,Gierz2017,Tielrooij2013,Mihnev2016}.
This separation of timescales makes graphene a paradigmatic platform for studying interaction-driven nonequilibrium dynamics.

From a theoretical point of view, however, this regime remains challenging. Widely used approaches such as semiclassical Boltzmann descriptions, semiconductor or graphene Bloch equations, and related second-order many-body truncations have been highly successful in describing many aspects of ultrafast relaxation \cite{RossiKuhn02,Winzer2010,Malic2011,Winzer2013_1,Malic2013,Malic2017}.
At the same time, these approaches rely on assumptions whose range of validity is difficult to assess in a controlled manner. In particular, they effectively presume that the early-time dynamics is dominated by
relatively low-order scattering processes and that higher-order many-body correlations do not qualitatively modify the relaxation pathway \cite{Winzer2010,Malic2013}. While this expectation is often physically
plausible, it is rarely benchmarked against accurate many-particle exact real-time dynamics in an interacting lattice model.

The purpose of this work is to provide such a benchmark in a controlled setting. We consider an interacting tight-binding description of finite graphene flakes and study the real-time evolution following a
simple particle--hole quench. The initial state is chosen as a non-interacting Slater determinant with a single particle promoted from the highest occupied level to an unoccupied level. This setup is minimal enough
to allow systematic comparison across different many-body truncation strategies, yet sufficiently nontrivial to generate interaction-induced relaxation and correlation growth during the subsequent time evolution.

Our central question is whether the post-quench dynamics can be captured within low-order particle--hole excitation spaces, or whether higher-order sectors become essential already on ultrafast timescales. To
address this, we compare the full Fock-space evolution with simulations restricted to subspaces containing at most a given number of particle--hole excitations. In addition, we monitor the single-particle orbital entropy
derived from the one-particle reduced density matrix (1RDM). This quantity vanishes for a single Slater determinant
and increases once the dynamics generates correlations beyond an effectively single-particle description
\cite{Eisert2010}. It therefore serves as a compact diagnostic for the breakdown of mean-field-like or low-order interaction pictures.

This perspective is motivated by two broader considerations. First, it provides a direct way to test whether low-order many-body descriptions are sufficient for interaction-driven relaxation in finite graphene systems.
Second, it identifies a class of nonequilibrium problems that are straightforward to initialize from non-interacting states but can develop substantial many-body complexity in time, making them natural benchmark candidates for future quantum-computing simulation approaches \cite{Georgescu2014,Altman2021,Fauseweh2024,Maskara2025}.

The main result of this work is that the convergence with excitation order can depend strongly on geometry and boundary conditions. For periodic flakes we looked at, the dynamics is reproduced remarkably well already within low-order excitation spaces, indicating that the relaxation is dominated by comparatively simple scattering processes. For hard-wall boundaries, by contrast, both the orbital entropy and the decay of the initial excitation reveal substantial contributions from higher-order excitation sectors. This demonstrates that apparent agreement at low excitation order is not universal and can break down in confined geometries even when the microscopic interaction remains unchanged.

The paper is organised as follows. In Sec.~\ref{sec:exp_overview} we briefly summarise the experimental status of ultrafast carrier relaxation in graphene and the characteristic timescales relevant for this work. In Sec.~\ref{sec:theory_overview} we review theoretical approaches to interaction-induced ultrafast dynamics and introduce the interacting lattice model and excitation-space construction used here. Section~\ref{sec:results} presents the real-time results and analyses the convergence with excitation order for different boundary conditions and system sizes. We conclude with a discussion of the implications for low-order many-body descriptions and for future simulation strategies.

\section{Experimental Overview}\label{sec:exp_overview}
The linear Dirac spectrum of graphene produces a two-dimensional electron fluid whose efficient Coulomb scattering scrambles energy
and momentum on ultrashort time scales, underpinning many prospective optoelectronic applications~\cite{Bonaccorso2010}.
While the equilibrium electronic structure of graphene is comparatively simple and well captured by a non-interacting tight-binding model on the honeycomb lattice, its ultrafast nonequilibrium dynamics is driven by electron--electron interaction that may exceed the validity of perturbative approaches.

A central experimental objective has therefore been to pin down the carrier–carrier thermalisation time, \(\tau_{ee}\),
defined as the interval over which an initially non-thermal distribution evolves into a hot Fermi–Dirac state.
Beginning with the 85 fs 780 nm pump–probe work of Dawlaty \textit{et al.}~\cite{Dawlaty2008},
which suggested a sub-100-fs decay of the photoinduced transmission in epitaxial graphene,
successive experiments have pushed the temporal resolution ever deeper into the few-femtosecond regime and have converged on a
remarkably consistent thermalisation window.  Sub-10-fs white-light spectroscopy by Breusing \textit{et al.}~\cite{Breusing2011}
revealed that a hot Fermi–Dirac distribution emerges within roughly 30 fs, a finding reinforced by the two-colour 7 fs/13 fs study of
Brida \textit{et al.}~\cite{Brida2013}.  Jensen \textit{et al.}~\cite{Jensen2014} expanded the pump–probe bandwidth and showed that this
timescale is essentially fluence-independent over an order of magnitude in carrier density.
Momentum-resolved snapshots obtained with extreme-UV time- and angle-resolved photoemission (tr-ARPES) by
Gierz and co-workers~\cite{Gierz2013,Gierz2015,Gierz2017} tracked the collapse of a non-thermal carrier
step into a hot Fermi sea in \(\sim\)25 fs, a result corroborated by Johannsen \textit{et al.}~\cite{Johannsen2013}
on quasi-free-standing monolayers.  In the THz domain, Tielrooij \textit{et al.}~\cite{Tielrooij2013} linked a \(50\pm10\) fs
electronic cascade to a transient sign flip in the photoconductivity, while Boltzmann-model fits to gate-tunable
measurements by Tomadin \textit{et al.}~\cite{Tomadin2018} and fluence-dependent studies by Mihnev \textit{et al.}~\cite{Mihnev2016}
confirmed that electronic thermalisation is complete in \(\lesssim\)30–50 fs irrespective of disorder or carrier density. 
Longer-time cooling experiments by George \textit{et al.}~\cite{George2008},
Strait \textit{et al.}~\cite{Strait2011} and König-Otto \textit{et al.}~\cite{KonigOtto2016} situate this ultrafast
process within a broader relaxation hierarchy, documenting optical-phonon emission on the 0.1–1 ps scale and acoustic or supercollision pathways on tens to hundreds of picoseconds, but they do not modify the initial carrier–carrier window. 
Taken together, these optical, tr-ARPES and THz studies—spanning pulse durations from 7 fs to 100 fs,
photon energies from THz to extreme-UV and a wide variety of sample preparations—cluster tightly around a carrier–carrier thermalisation
time of 25–35 fs, only rarely exceeding 50 fs even at low excitation densities; accordingly, we adopt \(\tau_{ee}\approx30~\text{fs}\)
as a realistic and experimentally validated benchmark for the theory presented below.

\section{Theoretical treatment of interaction–induced ultrafast carrier dynamics}
\label{sec:theory_overview}

Ultrafast optical or electrical pumps launch carrier populations far from equilibrium on sub-100 fs time scales;
inter-particle interactions then re-distribute energy and momentum and nudge the system toward a quasi-thermal state. 
Electron–electron (e–e) scattering dominates these earliest stages of decoherence and energy sharing, 
yet its microscopic description is notoriously difficult, so a hierarchy of complementary approaches has emerged.  
The most economical is the time-dependent Boltzmann equation, where momentum-resolved distribution functions are propagated with 
perturbative collision integrals for e–e, electron–phonon (e–ph) and impurity scattering; it reproduces carrier cooling and intervalley 
transfer at low excitation density but assumes instantaneous golden-rule rates, fixed quasiparticle energies and no quantum coherence, 
hence it breaks down below roughly 50 fs or at high fluence \cite{RossiKuhn02}.  Quantum-kinetic density-matrix theory and its 
semiconductor Bloch-equation realisation retain interband coherences, include phase-space filling and Hartree–Fock shifts, 
and add e–e scattering through a self-consistent second-Born self-energy; they capture Rabi cycling and excitonic transients 
but over-estimate thermalisation when dynamical screening or plasmon satellites become important \cite{RossiKuhn02}.  
Nonequilibrium Green’s-function methods on the Kadanoff–Baym/Keldysh contour offer a fully conserving and memory-preserving 
framework \cite{Keldysh65,Danielewicz1984}; with the generalised Kadanoff–Baym ansatz they provide single-time equations that 
embed spectral information, respect charge and energy conservation, and have revealed, for example, that population decay cannot 
be read directly from the imaginary part of the retarded self-energy and that Matthiessen’s rule fails when several channels 
compete \cite{Kemper2018}.  Their cubic scaling with simulation time and the need for dense $\mathbf k$ meshes currently limit 
applications to a few hundred femtoseconds.  First-principles real-time many-body perturbation theory augments density-functional 
band structures by \emph{ab-initio} e–e and e–ph matrix elements: either the rates feed a Boltzmann solver—efficient but typically 
employing static screening—or electrons and the screened Coulomb interaction are propagated self-consistently in real time 
within $GW\!+\,$Kadanoff–Baym, which is rigorous but restricted to small systems and short timescales \cite{Stefanucci2013,RungeGross84,Perfetto2015,Perfetto2018,Giustino2017}.  
For strongly correlated bands, nonequilibrium dynamical mean-field theory solves a self-consistent impurity model on the Keldysh contour, 
treating the local Hubbard interaction non-perturbatively and describing doublon production, 
Mott-gap collapse and pre-thermal plateaus \cite{Aoki2014,Eckstein2010}; 
however, it neglects non-local scattering and long-range Coulomb tails, so its predictive power is limited.
In one dimension, time-dependent density-matrix renormalisation group (t-DMRG) yields essentially exact
e–e dynamics \cite{DMRG_PS,DMRG_1,DMRG_2,DMRG_3,ejima2024}.  Entanglement growth still limits the
reachable times and prevents straightforward extension to the quasi-two-dimensional systems of current experimental interest.

Having outlined the general toolbox, we now focus on graphene where 
ultrafast e–e dynamics have been studied in exquisite detail.

\subsection{Ultrafast carrier dynamics in graphene}
\label{sec:theory_graphene}

Ultrafast carrier dynamics in graphene is widely treated using the graphene Bloch-equation (GBE) formalism,
a density-matrix theory that truncates the correlation hierarchy at the second-Born level and treats carrier–carrier scattering 
with statically or dynamically screened Coulomb matrix elements.  Early GBE calculations predicted efficient Auger-type carrier 
multiplication and thermalisation within 30–50 fs \cite{Winzer2010}.  
Subsequent studies added full pump–probe protocols and time-dependent screening \cite{Malic2011}, 
fluence-dependent scattering \cite{Winzer2013_1}, Landau quantisation \cite{Wendler2015}, and Fermi-level tuning via doping 
\cite{Kadi2015}.  Across the board, GBE simulations show that impact ionisation and Auger recombination dominate the first few tens of 
femtoseconds, producing a hot Fermi–Dirac distribution that cools through optical phonons on sub-picosecond time 
scales \cite{Brida2013,Kadi2014,Mihnev2016}.  Because the formalism keeps phase coherence and two-particle correlations explicitly, 
it can reproduce subtle effects such as transient population inversion and optical gain \cite{Winzer2013_2}, 
negative differential transmission \cite{Kadi2014}, and the recently observed double-bended bleaching \cite{Winzer2017}.  
Comprehensive reviews are given in Refs.~\cite{Malic2013,Malic2017,Winnerl2017}.

Where phase coherence is less critical—e.g.\ when extracting effective scattering rates or modelling high-fluence relaxation—the 
photo-excited population is often treated with a semi-classical Boltzmann-transport equation that contains statically or dynamically 
screened Coulomb kernels.  Using a static-RPA kernel, Butscher \emph{et al.} obtained thermalisation times of 50–100 fs \cite{Butscher2007}, 
while the inclusion of dynamical screening reduced this to 20–30 fs \cite{Kim2011,Hwang2007}.  
The favourable scaling of Boltzmann solvers makes them attractive for systematic sweeps over fluence, 
temperature or substrate permittivity and for benchmarking the full many-body GBE codes, as demonstrated in Refs.~\cite{Kim2011,Song2013}. 
Since the ultra-fast relaxation is purely interaction driven we want to address the question,
whether resting the dynamics to second order approximations is a justified approximation.

\subsection{Dynamics of an interacting tight-binding model}
\label{sec:tight_binding_intro}

\begin{figure}
    \center
    \includegraphics[width=0.45\textwidth]{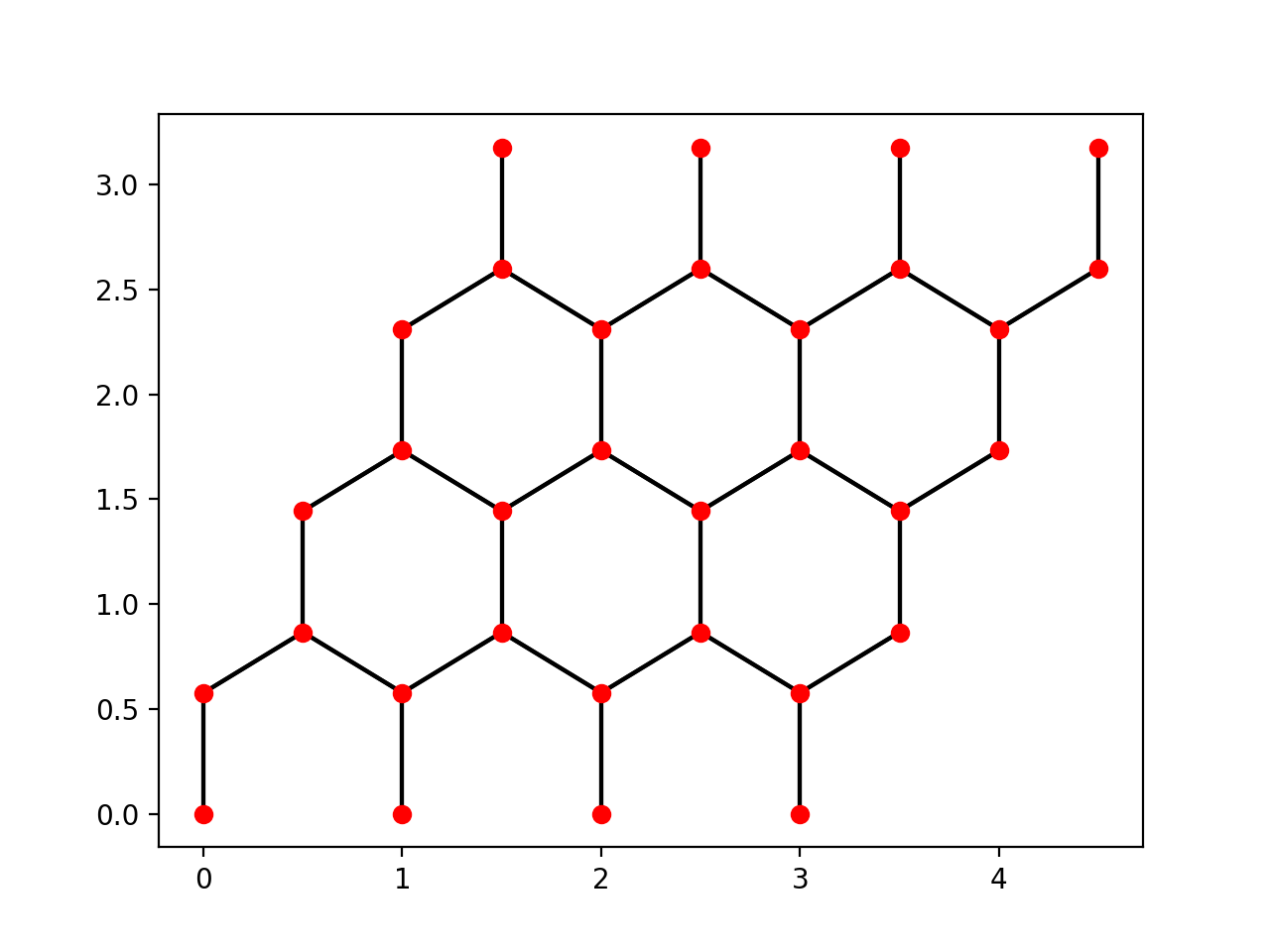}
    \caption{Schematic representation of the $4\times4$ graphene flake used in this work, comprising 32 lattice sites on the honeycomb lattice. The units are in lattice constants.}
\label{fig:1a}
\end{figure}

To address the above question, we study interacting lattice models for finite graphene flakes at half filling. 
The full model can be formulated for both spinless and spinful fermions on the honeycomb lattice with local and non-local density--density interactions. 
In the present work, however, we restrict ourselves to the spinless case,
\begin{equation}
  \HH
  = \sum_{ij} t_{ij}\,\hat c_i^\dagger \hat c_j
  + \sum_{ij} U_{ij}
    \left(\hat n_i - \frac{1}{2}\right)
    \left(\hat n_j - \frac{1}{2}\right),
  \label{eq:H_realspace}
\end{equation}
where $t_{ij}$ denotes the hopping amplitude between real-space orbitals $i$ and $j$, $U_{ij}$ the interaction matrix element, and $\hat n_i=\hat c_i^\dagger \hat c_i$.

In the following, we consider a $4\times4$ graphene flake comprising 32 lattice sites, as illustrated in Fig.~\ref{fig:1a}, with either periodic or hard-wall boundary conditions. This number follows from the fact that graphene has a two-atom basis (A and B sublattices), so a $4\times4$ array of unit cells contains $4\times4\times2 = 32$ atomic sites.

A notable feature of the graphene lattice is that, even for comparatively strong interactions, the one-particle reduced density matrix of the ground state
\begin{equation}
  \rho_{ij}
  = \bra{\Psi_0}\hat c_i^\dagger \hat c_j\ket{\Psi_0}
  \label{eq:1rdm}
\end{equation}
often remains close to idempotent, i.e.\ its eigenvalues stay near $0$ or $1$. 
This indicates that the many-body state remains close to a single Slater determinant, or at least to a weakly correlated mean-field-like state. 
In addition, because the interaction is written in particle--hole-symmetric form, the Hartree contribution vanishes at half filling.

To mimic the effect of an optical pump in a minimal way, we initialize the system in the non-interacting ground state constructed from the single-particle eigenstates $u_{iq}$ of the hopping Hamiltonian,
\begin{align}
  \ket{\Psi_0}
  &= \prod_{q\le \kF} \hat c_q^\dagger \ket{0},
  \label{eq:GS_nonint}
  \\
  \hat c_q^\dagger
  &= \sum_i u_{iq}\,\hat c_i^\dagger .
  \label{eq:cq_def}
\end{align}
We then create a single particle--hole excitation by removing one fermion from the highest occupied level $\kF$ and promoting it to an unoccupied level $\tilde p$,
\begin{equation}
  \ket{\xi_0}
  = \hat c_{\tilde p}^\dagger \hat c_{\kF}\ket{\Psi_0},
  \label{eq:xi0}
\end{equation}
where in the calculations presented below $\tilde p$ is chosen as the second-highest unoccupied level.

The subsequent real-time evolution is generated by the full interacting Hamiltonian,
\begin{equation}
  \ket{\xi(t)}
  = \e^{-\im(\HH-E_0)t}\ket{\xi_0},
  \label{eq:time_evolution}
\end{equation}
where $E_0$ is the ground-state energy.

The occupations of the non-interacting single-particle eigenlevels are obtained from the 1RDM by transforming Eq.~\eqref{eq:1rdm} from the real-space basis to the eigenbasis of the hopping Hamiltonian.

\subsubsection{Performing the time evolution}
Since the time evolution is governed by a time-independent Hamiltonian $\HH$ \eqref{eq:time_evolution}
we perform the time evolution by applying the matrix exponential for a given set of time points $\{ t_j \}$:
\begin{align}
  \ket{\xi(t_j)} &= \e^{-\im \left(\HH - E_0\right) \left(t_j - t_{j-1} \right)} \ket{\xi_{t_{j-1}}} \,,
\end{align}
where the the ground state energy $E_0$ is subtracted to avoid unnecessarily strong phase oscillations.
By using the matrix exponential we avoid any Trotter like errors and the time stepping for the time steps $\{ t_j \}$
can be set to any desired value. Typically we use a time stepping of the order of 0.2 in units of the hopping element.
For small Hibert spaces we calculated the full matrix exponential from the eigen decomposition of the matrix representation of $\HH$,
while for the largest Hilbert spaces we resort to a Krylov \cite{Krylov1931} subspace matrix exponential \cite{Saad:1992,DMRG_PS}.

We extract the time dependent density matrix \eqref{eq:1rdm} in a two step process, as this allows for an efficient
parallelization which is necessary for the larger systems, especially the 32 site flake.

\begin{align}
  \ket{\xi_j} &= \hat{c}_j \ket{\xi(t)}
  \\ \rho_{ij}(t)  &= \braket{\xi_i}{\xi_j}
\end{align}

\subsubsection{Single-particle orbital entropy}
Within any effective single-particle description, such as Hartree--Fock or time-dependent density-functional theory, the 1RDM evolves unitarily as
\begin{equation}
  \rho(t) = U(t)\,\rho(0)\,U^\dagger(t),
  \label{eq:rho_unitary}
\end{equation}
so its eigenvalues are constants of motion. Consequently, such a description cannot by itself generate relaxation of the natural occupations. In translationally invariant systems, this also implies the absence of any decay in momentum occupations. Strictily speaking we have to track the eigen basis with time. However, we can stick to the intial basis as the diagonalization is governed by the Fourier (PBC) / sine (hard-wall BC) transformation. We have checked that looking at the eigen values by re-diagonalizing $\rho(t)$ at each time steps does not alter our results.

To quantify the build-up of correlations during the interacting time evolution, we therefore consider the single-particle orbital entropy associated with the 1RDM,
\begin{align}
  \Ssp
  &= - \Tr\!\left[\rho \ln \rho\right]
     - \Tr\!\left[(1-\rho)\ln(1-\rho)\right]
  \nonumber\\
  &= - \sum_\ell \Bigl[
      n_\ell \ln n_\ell
      + (1-n_\ell)\ln(1-n_\ell)
     \Bigr],
  \label{eq:OrbitalEntropy}
\end{align}
where $n_\ell$ are the eigenvalues of $\rho$, i.e.\ the natural occupations of the single-particle orbitals \cite{PeschelEisler:2009}.

For a pure Slater determinant, such as the initial non-interacting ground state $\ket{\Psi_0}$, all natural occupations are exactly $0$ or $1$, and therefore $\Ssp=0$. A finite value of $\Ssp$ directly signals that the interacting dynamics has generated correlations beyond an effectively single-particle description.

We emphasize that, unlike bipartite entanglement entropies, the orbital entropy does not depend on an arbitrary spatial partition of the system. Instead, it depends only on the spectrum of the 1RDM and is therefore basis independent at the one-particle level.

\subsubsection{Restriction to particle--hole excitation spaces}
In addition to the full time evolution, we consider a controlled Hilbert-space truncation in the basis of non-interacting single-particle eigenstates. In that basis, the Hamiltonian takes the form
\begin{equation}
  \HH
  = \sum_\ell \varepsilon_\ell \hat c_\ell^\dagger \hat c_\ell
  + \sum_{p\ell mq}
    U_{p\ell mq}\,
    \hat c_p^\dagger \hat c_\ell^\dagger \hat c_m \hat c_q,
  \label{eq:H_eigenbasis}
\end{equation}
where $\varepsilon_\ell$ are the single-particle energies and $U_{p\ell mq}$ the interaction matrix elements in that basis.

Although exact simulations in this representation are generally more expensive than in real space, it provides a natural framework for constructing systematically truncated excitation spaces. We start from the one-dimensional reference space
\begin{equation}
  \VV_0 = \Span\{\ket{\xi_0}\}.
  \label{eq:single_state_V0}
\end{equation}

We then define a particle--hole excitation operator that promotes a fermion from an occupied level $h$ below the Fermi energy to an unoccupied level $p$ above it,
\begin{equation}
  \hat P
  = \sum_{p,h} \hat c_p^\dagger \hat c_h .
  \label{eq:P_operator}
\end{equation}
Using this operator recursively, we construct the truncated excitation spaces
\begin{align}
  \VV_1 &= \Span\!\left\{\VV_0,\hat P\,\VV_0\right\},
  \label{eq:V1}
  \\
  \VV_n &= \Span\!\left\{\VV_{n-1},\hat P\,\VV_{n-1}\right\}.
  \label{eq:Vn_recursion}
\end{align}
By construction, $\VV_n$ contains all states reachable from the initial state by at most $n$ successive particle--hole excitations.

If excitation sectors with $n>2$ become quantitatively important, this indicates that the dynamics cannot be captured within a simple low-order picture based only on one- and two-particle--hole processes. This provides a practical diagnostic for the breakdown of low-order approximations. We stress, however, that the converse is not strictly true: even if the dynamics is well approximated within $\VV_{\le 2}$, this does not imply exact equivalence to any specific second-order diagrammatic scheme, since the truncated space still includes repeated couplings and therefore partial infinite resummations within that subspace.

\section{Results}
\label{sec:results}

We now turn to the central objective of this work, namely to assess whether interaction-driven relaxation following a particle--hole quench in graphene flakes can be reliably captured within low-order excitation spaces. This question is directly relevant for widely used second-order many-body approaches such as the graphene Bloch equations, which implicitly assume that relaxation dynamics are dominated by processes involving only a small number of particle--hole excitations. To provide a controlled diagnostic of this assumption, we compare the full Fock-space dynamics with simulations restricted to subspaces containing up to $n$ particle--hole excitations.

In the following, we consider the spinless interacting graphene model defined in Eq.~\eqref{eq:H_realspace} with density--density interactions up to third neighbors, with interaction strengths $U_1=0.2$, $U_2=0.1$, and $U_3=0.05$, where $U_1$, $U_2$, and $U_3$ denote nearest-, next-nearest-, and next-next-nearest-neighbor interactions, respectively. All energies are measured in units of the nearest-neighbor hopping amplitude $t$.
We emphasize, however, that the qualitative conclusions do not rely on this particular parameter choice. In Appendix~\ref{app:u_scan_graphene}, we present results for a range of nearest-neighbor interaction strengths $U_1$ with $U_2=U_3=0$, demonstrating that the observed behavior persists beyond the parameter set used in the main text. 
Furthermore, in Appendix~\ref{app:bandstructure}, we show the interacting band structure obtained from cluster perturbation theory \cite{enenkel2023, Senechal2000}. This confirms that, for the interaction strengths considered here, the Dirac cone remains intact and no interaction-induced gap opens.

\begin{figure}
  \center
    \includegraphics[width=0.98\textwidth]{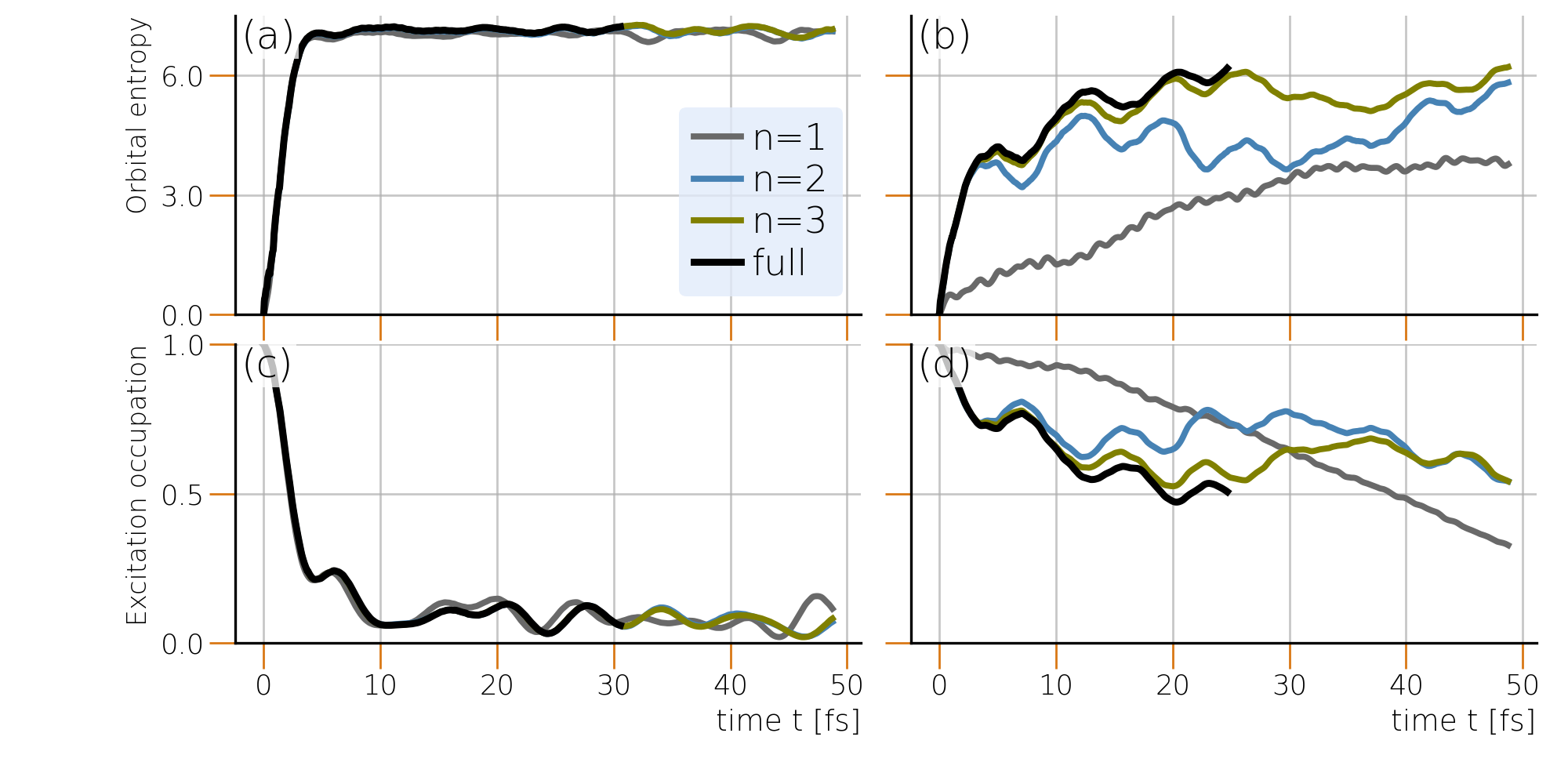}
    \caption{
Time evolution after a particle--hole quench in a $4\times4$ graphene flake (32 sites). 
The upper panels show the orbital entropy $S_\mathrm{sp}$, while the lower panels display the occupation of the initially excited single-particle level $n_{\tilde{p}}$. 
The left panels correspond to periodic boundary conditions and the right panels to hard-wall boundary conditions. 
Colored curves denote simulations restricted to at most $n=1$ (grey), $n=2$ (blue), and $n=3$ (olive) particle--hole excitations, while black curves represent the full Fock-space dynamics. 
Periodic systems are already well described by low-order excitation spaces ($n\le2$), whereas hard-wall systems require higher-order excitations and exhibit slower relaxation.
}
\label{fig:1}
\end{figure}

Figure~\ref{fig:1} shows the time evolution following the quench in a $4\times4$ graphene flake (32 sites), where one particle is removed from the highest occupied level (single-particle index $15$) and promoted to a higher-energy level (index $30$). 
We focus on two complementary observables. First, the orbital entropy $\Ssp$ defined in Eq.~\eqref{eq:OrbitalEntropy} quantifies the deviation of the 1RDM from an idempotent form and therefore measures the buildup of correlations. Second, the occupation of the initially excited level $n_{\tilde{p}}$ provides a direct probe of relaxation dynamics.

For periodic boundary conditions, the dynamics is characterized by a rapid growth of orbital entropy accompanied by a fast decay of the excited-level occupation on sub-10~fs timescales. 
The comparison between excitation spaces reveals only a weak hierarchy. Even restricting the Hilbert space to a single particle--hole excitation already captures the qualitative behavior of the decay, with only small quantitative deviations from the full Fock-space evolution. Including up to two excitations yields results that are nearly indistinguishable from the exact dynamics, while the inclusion of third-order excitations produces only negligible corrections. These observations indicate that, in the periodic system, relaxation is dominated by low-order scattering processes and is well described already at very low excitation order.

A qualitatively different and somewhat counterintuitive picture emerges for hard-wall boundary conditions. While the breaking of translational invariance eliminates momentum conservation and thus increases the nominal scattering phase space, both entropy growth and occupation decay are substantially slower, occurring on timescales exceeding $30$~fs. 
More importantly, the hierarchy of excitation spaces does not exhibit convergence at second order: the dynamics obtained within the $n=2$ excitation space differs markedly from both the $n=3$ and the full Fock-space results, demonstrating that higher-order correlations contribute significantly to the relaxation process.

\begin{figure}
    \includegraphics[width=0.48\textwidth]{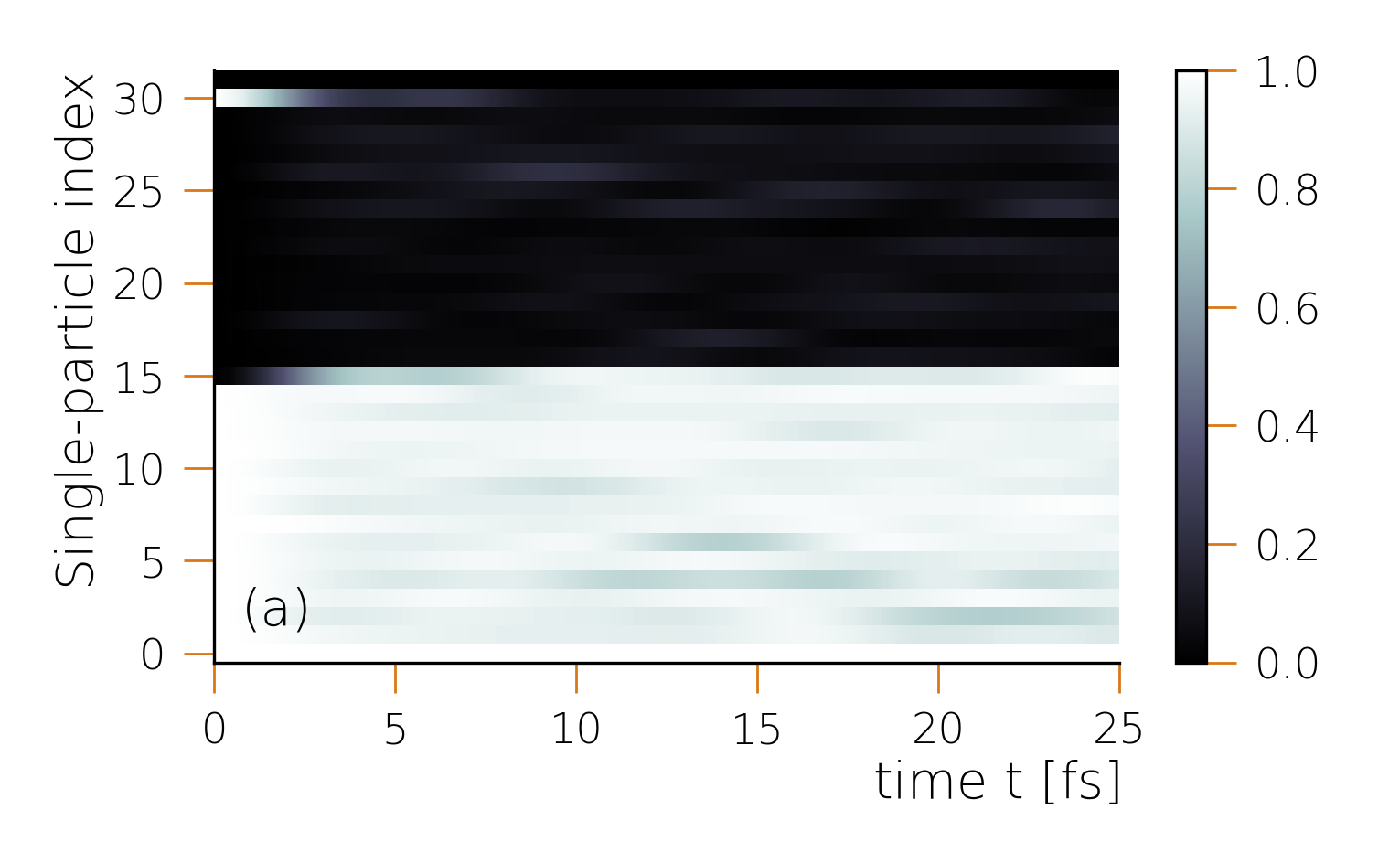}
    \includegraphics[width=0.48\textwidth]{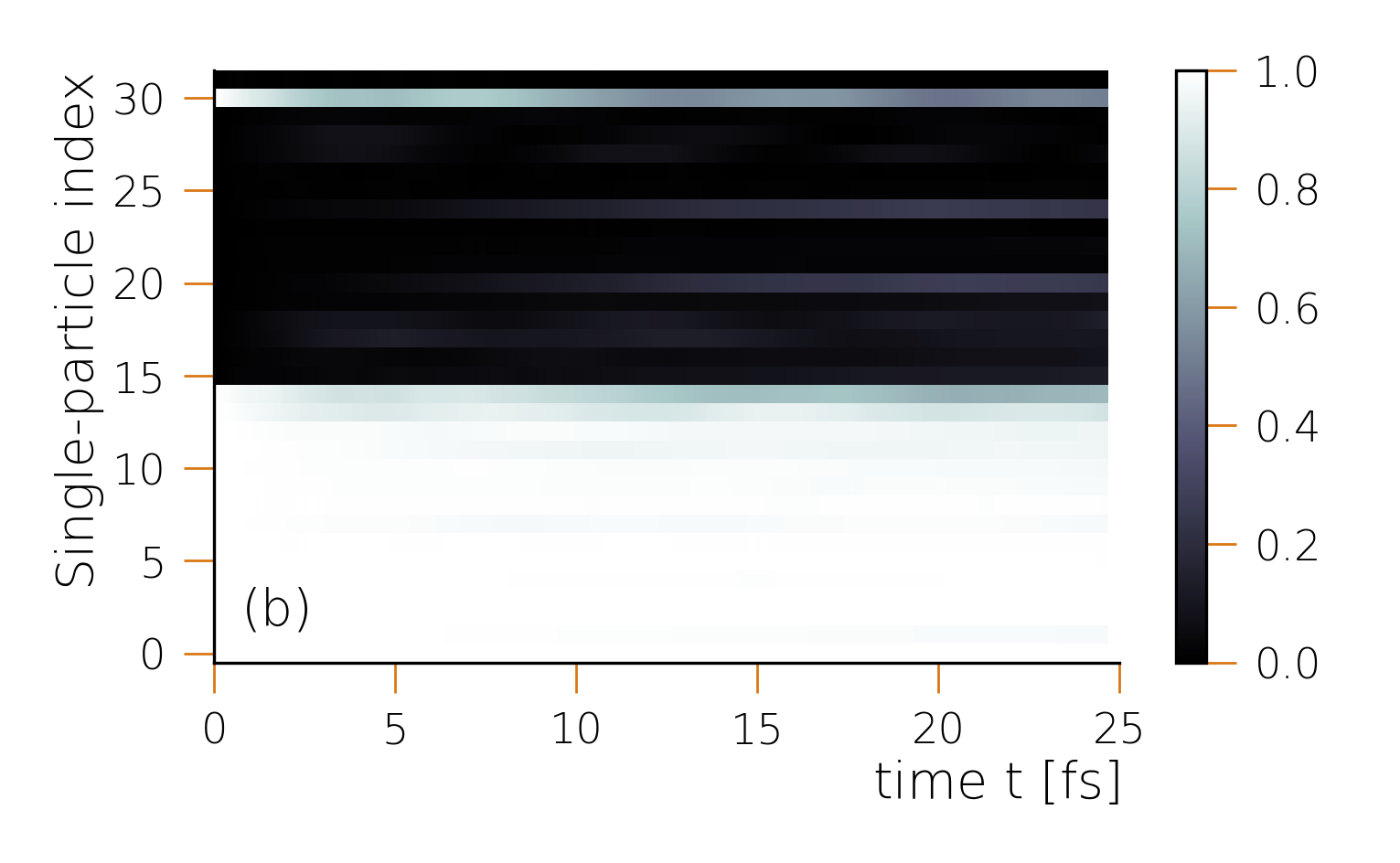}
\caption{
Full Fock-space evolution of all single-particle occupations $n_q$ following the quench in a $4\times4$ graphene flake. 
Single-particle indices on the vertical axis are ordered by the non-interacting eigenenergies $\epsilon_\ell$. 
White (black) indicates occupation one (zero). 
Panel (a) shows periodic boundary conditions and panel (b) hard-wall boundary conditions. 
Periodic systems show rapid and efficient redistribution of the excitation, while hard-wall systems display slower and more structured dynamics indicative of higher-order processes.
}
\label{fig:2}
\end{figure}

The microscopic redistribution of single-particle occupations is illustrated in Fig.~\ref{fig:2}, which shows the full Fock-space evolution of the occupation spectrum in the eigenbasis of the non-interacting Hamiltonian. The decay of the initially excited level (index $30$), shown in Fig.~\ref{fig:1}(c,d), corresponds to the horizontal evolution of this level in Fig.~\ref{fig:2}. 

For periodic boundary conditions, scattering rapidly redistributes the initial excitation, consistent with fast thermalisation. Already after $\sim 5$~fs, the initially excited level at index $30$ is nearly empty, while the source level at index $15$ is nearly refilled. 
In contrast, hard-wall boundaries lead to a slower and more structured redistribution pattern, suggesting that relaxation proceeds through collective multi-particle processes rather than simple pair scattering.

\begin{figure*}[!t]
    \centering
    \includegraphics[width=0.95\linewidth]{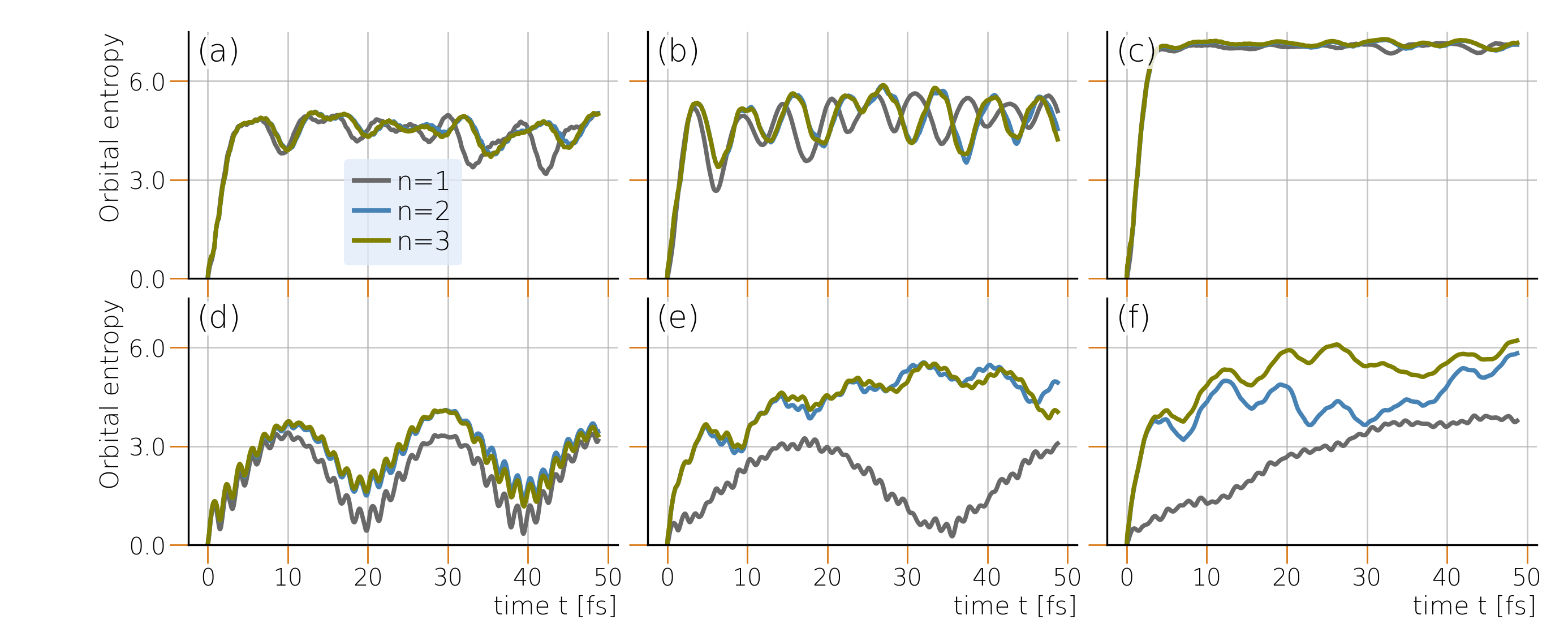}
\caption{
System-size dependence of excitation-space convergence. 
Time evolution of the orbital entropy for periodic (a--c) and hard-wall (d--f) boundary conditions for systems of increasing size: $4\times2$ (a,d), $4\times3$ (b,e), and $4\times4$ (c,f). 
Grey, blue, and olive curves correspond to simulations restricted to $n=1$, $n=2$, and $n=3$ particle--hole excitations, respectively. 
Low-order excitation spaces remain accurate for periodic graphene across system sizes, whereas hard-wall systems show increasing deviations and growing importance of higher-order correlations.
}
\label{fig:3}
\end{figure*}

In Fig.~\ref{fig:3} we analyse systems of increasing size, comparing the orbital entropy dynamics obtained within the $n=1$, $n=2$, and $n=3$ excitation spaces. 
For periodic systems, the differences between these excitation sectors remain small across all sizes considered, supporting the conclusion that low-order processes dominate relaxation in translationally invariant geometries. 
In contrast, hard-wall systems display a pronounced size dependence. While smaller flakes may suggest apparent convergence already at $n=2$, larger systems exhibit substantial deviations between successive excitation sectors, signalling the growing importance of higher-order correlations. This demonstrates that conclusions drawn from small-system calculations can be misleading when extrapolated toward experimentally relevant system sizes.

In Appendix~\ref{app:u_scan_graphene}, we present results for the $4\times4$ system with varying nearest-neighbor interaction strength $U_1$ and $U_2=U_3=0$. In Appendix~\ref{app:u_scan_squarelattice}, we repeat this parameter scan for a $6\times5$ square lattice. 
While all systems show the expected trend that higher-order processes become more relevant with increasing interaction strength, graphene flakes with periodic boundary conditions represent a special case where low-order truncations remain quantitatively accurate. In contrast, for hard-wall graphene flakes and for the square lattice, the difference between low- and higher-order excitation spaces is more pronounced, and boundary-condition effects are considerably weaker in the latter.

Taken together, these results establish a practical diagnostic criterion for the applicability of low-order many-body approaches. Rapid convergence with excitation order indicates that relaxation is dominated by low-order scattering processes and that perturbative descriptions are adequate. Conversely, persistent discrepancies between successive excitation sectors signal regimes in which higher-order correlations play a decisive role and low-order truncations become unreliable. 

The latter situation is particularly relevant for confined graphene systems and provides a concrete example of interaction-driven dynamics that is straightforward to initialize from a non-interacting state yet generates substantial many-body correlations during time evolution. This combination makes the quench protocol studied here a natural benchmark for future quantum-simulation approaches, where efficient state preparation can be combined with intrinsically many-body real-time dynamics beyond the reach of classical perturbative methods.

\section{Summary and discussion}
\label{sec:summary}

In this work we studied interaction-driven ultrafast dynamics in finite graphene flakes by means of a controlled
particle--hole quench. Starting from a non-interacting Slater-determinant ground state with one particle promoted
across the Fermi level, we compared the exact real-time evolution in full Fock space with dynamics restricted to subspaces containing at most a given number of particle--hole excitations. This construction provides a direct diagnostic for the extent to which the relaxation can be understood in terms of low-order many-body processes.

Our analysis focused on two complementary observables: the occupation dynamics of the initially excited single-particle level and the orbital entropy associated with the one-particle reduced density matrix. 
The latter is particularly useful because it quantifies the departure from an idempotent one-particle reduced density matrix and therefore tracks the buildup of correlations beyond an effectively single-particle description. In this sense, it offers a compact and basis-independent measure of correlation growth during the nonequilibrium evolution.

The results reveal a clear and physically relevant dichotomy. For periodic boundary conditions, the quench dynamics
is reproduced very well already within low-order excitation spaces. In particular, the inclusion
of up to two particle--hole excitations captures both the rapid decay of the initially excited level and the
corresponding growth of orbital entropy with only minor quantitative deviations from the full many-body evolution. Within the present setup, this indicates that the dominant relaxation channels can indeed be described in terms of comparatively low-order scattering processes.

For hard-wall boundary conditions, the picture changes qualitatively. Although translational invariance is broken
and momentum conservation is relaxed, the dynamics becomes slower and the convergence with excitation order deteriorates substantially. In this case, restricting the evolution to low-order particle--hole sectors does not
provide a quantitatively reliable description, and higher-order excitation spaces contribute significantly on the timescales of interest. This demonstrates that the effectiveness of low-order truncation schemes is strongly geometry dependent and cannot be inferred solely from the microscopic interaction strength.

A central message of our work is methodological: convergence with particle--hole excitation order provides a practical benchmark for the applicability of low-order many-body descriptions. Rapid convergence suggests that
perturbative or low-order correlation treatments are sufficient to capture the essential physics of the quench. Persistent discrepancies between successive excitation sectors, by contrast, signal that higher-order correlations
are dynamically relevant and that low-order approximations must be treated with caution. Importantly, this criterion does not establish a strict one-to-one correspondence with any particular diagrammatic approximation, such as
second-Born schemes, but it does offer a controlled way to diagnose whether the underlying dynamics is dominated by a small or large hierarchy of correlated processes.

The present results also raise an interesting physical question: why do confined flakes with hard-wall boundaries exhibit slower relaxation and stronger higher-order contributions than their periodic counterparts? Our data suggest
that the answer is not simply an increase or decrease in the nominal scattering phase space. Instead, the boundary conditions modify the structure of the single-particle eigenstates and the many-body spectrum, thereby reshaping how
the initial quench couples to correlated excitation pathways. 
We have checked that adding weak or moderate disorder to system with PBC doesn't lead to such an increase of the interaction effect. We also checked, that this fundamental difference between PBC and HWBC exists in flakes where no edge states are present. A more detailed analysis of this mechanism, for example in terms of level spacings, edge-state character, or matrix-element structure, would be a valuable subject for future work.

More broadly, the quench protocol studied here defines a useful class of benchmark problems for
nonequilibrium many-body simulation. The initial state is easy to prepare because it is generated from a non-interacting reference state by a simple particle--hole excitation, while the subsequent time evolution can nevertheless develop substantial correlations and sensitivity to boundary conditions. This combination makes the problem attractive not only for testing classical approximation schemes but also for assessing future quantum-computing simulation strategies, where simple state preparation and intrinsically many-body real-time dynamics form a particularly natural match.

Several extensions of the present study are possible. On the modelling side, it will be important to include spinful fermions, larger flakes, and interaction profiles closer to realistic screened Coulomb matrix elements in graphene. On the methodological side, a more explicit comparison to established second-order kinetic approaches would help to translate the present excitation-space benchmark into the language of commonly used ultrafast theories. Such developments would clarify to what extent the boundary-condition effects identified here survive in larger and more realistic systems.

In summary, we have shown that a particle--hole quench in interacting graphene flakes provides a controlled probe of ultrafast correlation buildup and a stringent test for low-order many-body descriptions.
While periodic systems are captured surprisingly well by low-order excitation spaces, confined systems with hard-wall boundaries exhibit substantial higher-order contributions already on ultrafast timescales. The orbital entropy offers a simple and effective marker for this breakdown of low-order descriptions.
Our results therefore identify both the promise and the limitations of reduced many-body treatments of interaction-driven relaxation in finite graphene systems.



\paragraph{Funding information}
This project was made possible by the DLR Quantum Computing Initiative and the Federal Ministry for Research, Technology and Space (\href{qci.dlr.de/projects/ALQU}{\sf qci.dlr.de/projects/ALQU})

\begin{appendix}
\numberwithin{equation}{section}

\section{Interaction dependence of excitation-space convergence: graphene flake}
\label{app:u_scan_graphene}

\begin{figure*}[!t]
    \centering
    \includegraphics[width=0.99\linewidth]{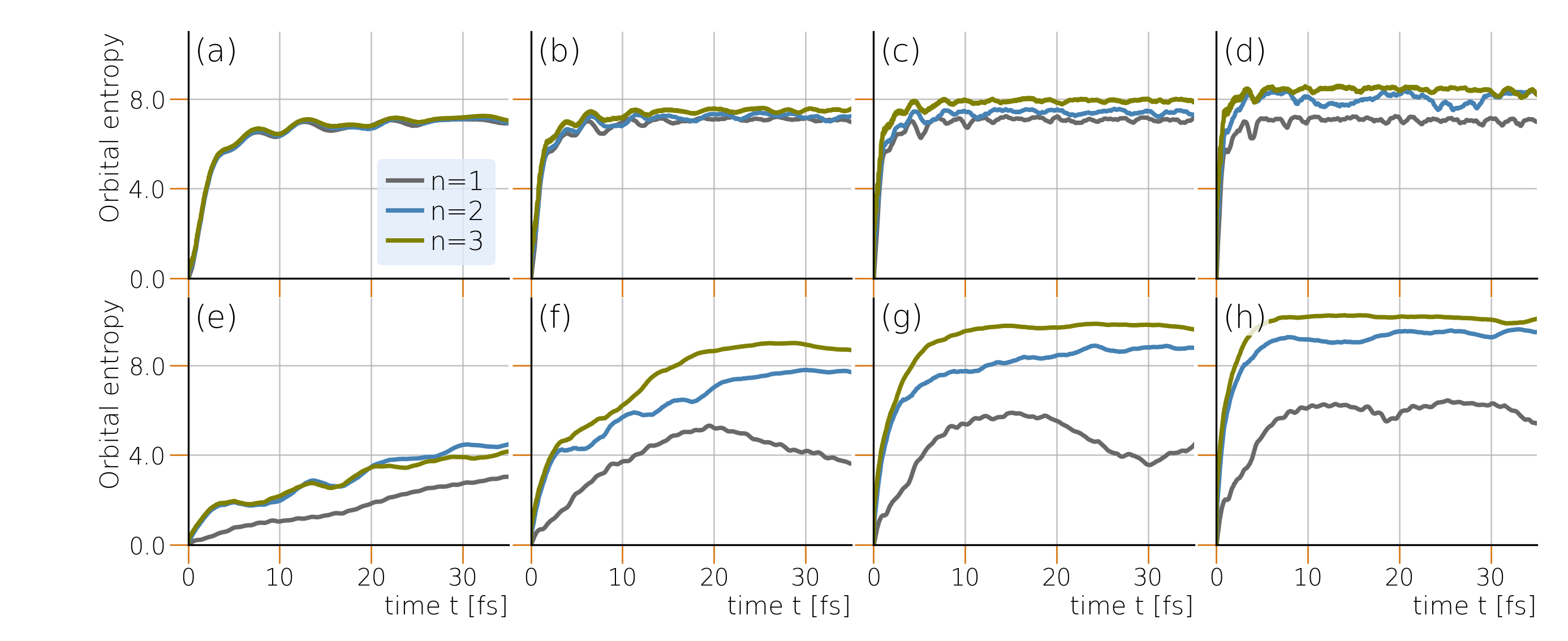}
\caption{
Interaction dependence of excitation-space convergence for a $4\times4$ graphene flake. 
Time evolution of the orbital entropy is shown for periodic (a--d) and hard-wall (e--h) boundary conditions at increasing nearest-neighbor interaction strength $U_1=0.2$, $0.4$, $0.6$, and $0.8$ (from left to right), with $U_2=U_3=0$. 
Grey, blue, and olive curves correspond to simulations restricted to $n=1$, $n=2$, and $n=3$ particle--hole excitation sectors, respectively. 
Increasing interaction strength enhances higher-order contributions in all cases, but periodic systems retain near-convergence at low excitation order over a much wider parameter range than hard-wall systems.
}
    \label{fig:A1}
\end{figure*}

In this appendix we analyze how the convergence with particle--hole excitation order evolves as a function of the interaction strength $U_1=0.2$, $0.4$, $0.6$ and $0.8$ for a fixed $4\times4$ graphene flake with $U_2=U_3=0$. The corresponding orbital-entropy dynamics is shown in Fig.~\ref{fig:A1}.

As expected, a general trend is that increasing interaction strength systematically enhances the importance of higher-order excitation sectors. This is reflected in the growing separation between results obtained in different excitation subspaces.

For periodic boundary conditions, the convergence with excitation order remains relatively robust at weak interaction. In particular, for $U_1=0.2$ in panel (a) the $n=1, n=2$ and $n=3$ results are nearly indistinguishable, indicating that low-order processes provide an accurate description of the dynamics. However, with increasing $U_1$ (b-d) the difference between excitation sectors grows steadily, demonstrating that higher-order correlations become progressively more relevant. This evolution is gradual, suggesting that the dominant relaxation channels remain low-order but acquire increasing higher-order corrections.

For hard-wall boundary conditions, the situation is qualitatively different. Already at moderate interaction strength, $U_1=0.4$, a pronounced deviation between the $n=1, n=2$ and $n=3$ results emerges. With further increasing $U_1$, this discrepancy becomes larger, indicating that higher-order excitation sectors are essential for a quantitative description of the dynamics.

\section{Interaction dependence on a square lattice}
\label{app:u_scan_squarelattice}

\begin{figure*}[!t]
    \centering
    \includegraphics[width=0.99\linewidth]{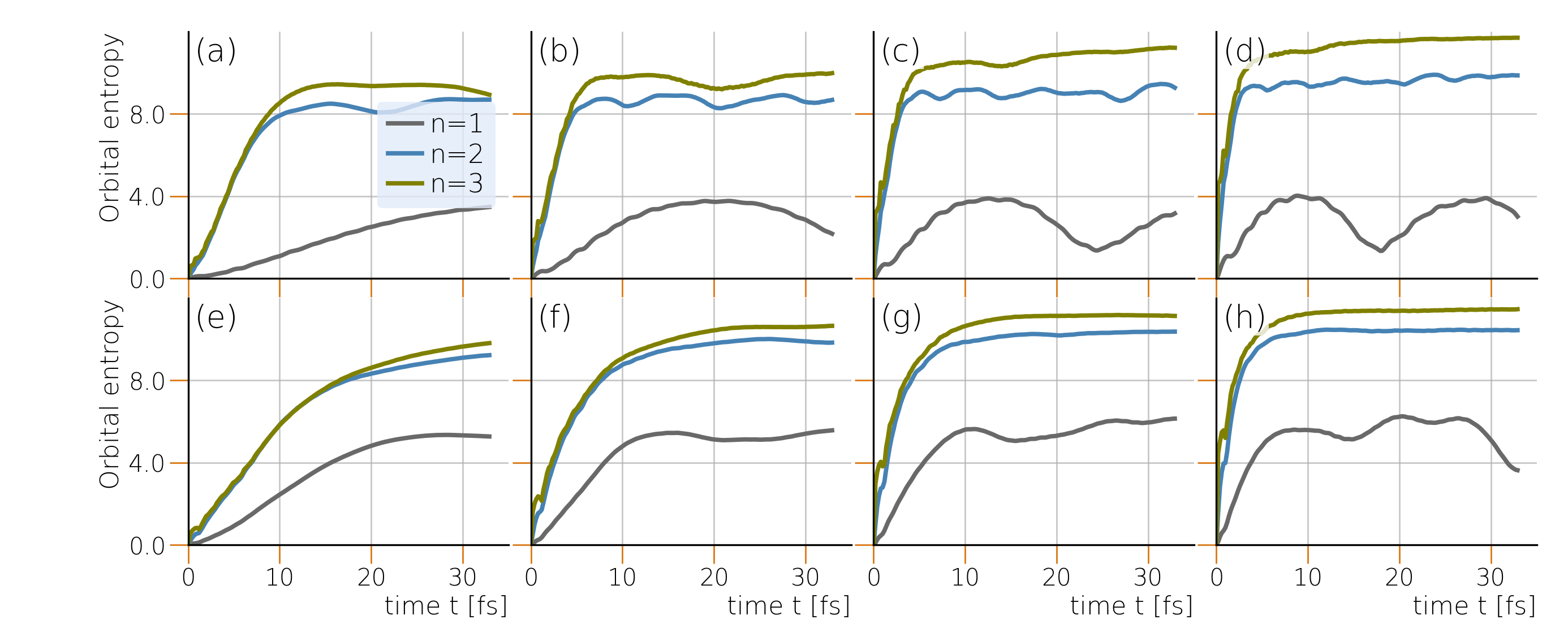}
\caption{
Interaction dependence of excitation-space convergence for a $6\times5$ square lattice. 
We display the same quantities and parameters as in Fig.~\ref{fig:A1}. 
Unlike graphene, the square lattice shows weak boundary-condition dependence and consistently requires higher-order excitation sectors, highlighting that the exceptional low-order behavior of periodic graphene is not generic.
}
    \label{fig:A2}
\end{figure*}

To assess the generality of the observations made for graphene, we repeat the interaction-strength analysis for a $6\times5$ square lattice. The corresponding results are shown in Fig.~\ref{fig:A2}.

In contrast to the graphene flake, the dependence on boundary conditions is much weaker in this case: periodic and hard-wall systems exhibit very similar behavior across the full interaction range. A clear hierarchy of excitation sectors is nevertheless present. The $n=1$ truncation significantly underestimates the entropy growth for all interaction strengths, demonstrating that single particle--hole excitations are insufficient even at weak coupling. Including $n=2$ excitations yields a substantial improvement, while the difference between $n=2$ and $n=3$ increases with increasing interaction strength.

Interestingly, the square lattice behaves more similar to the hard-wall graphene case than to periodic graphene. In both systems, higher-order excitation sectors contribute more visibly, and the convergence with excitation order is less pronounced. This indicates that the particularly good agreement at low excitation order observed for periodic graphene is not generic, but instead depends sensitively on the geometry.

Therefore, the comparison between Figs.~\ref{fig:A1} and \ref{fig:A2} highlights that the strong boundary-condition dependence found for graphene is not a universal feature of interacting lattice systems. Rather, it originates from the specific spectral and geometric properties of the honeycomb lattice. While increasing interaction strength generally enhances the role of higher-order correlations, the rate at which low-order descriptions break down is strongly system dependent.

\section{Band structure of the interacting graphene flake}
\label{app:bandstructure}

\begin{figure}
  \center
    \includegraphics[width=0.45\textwidth]{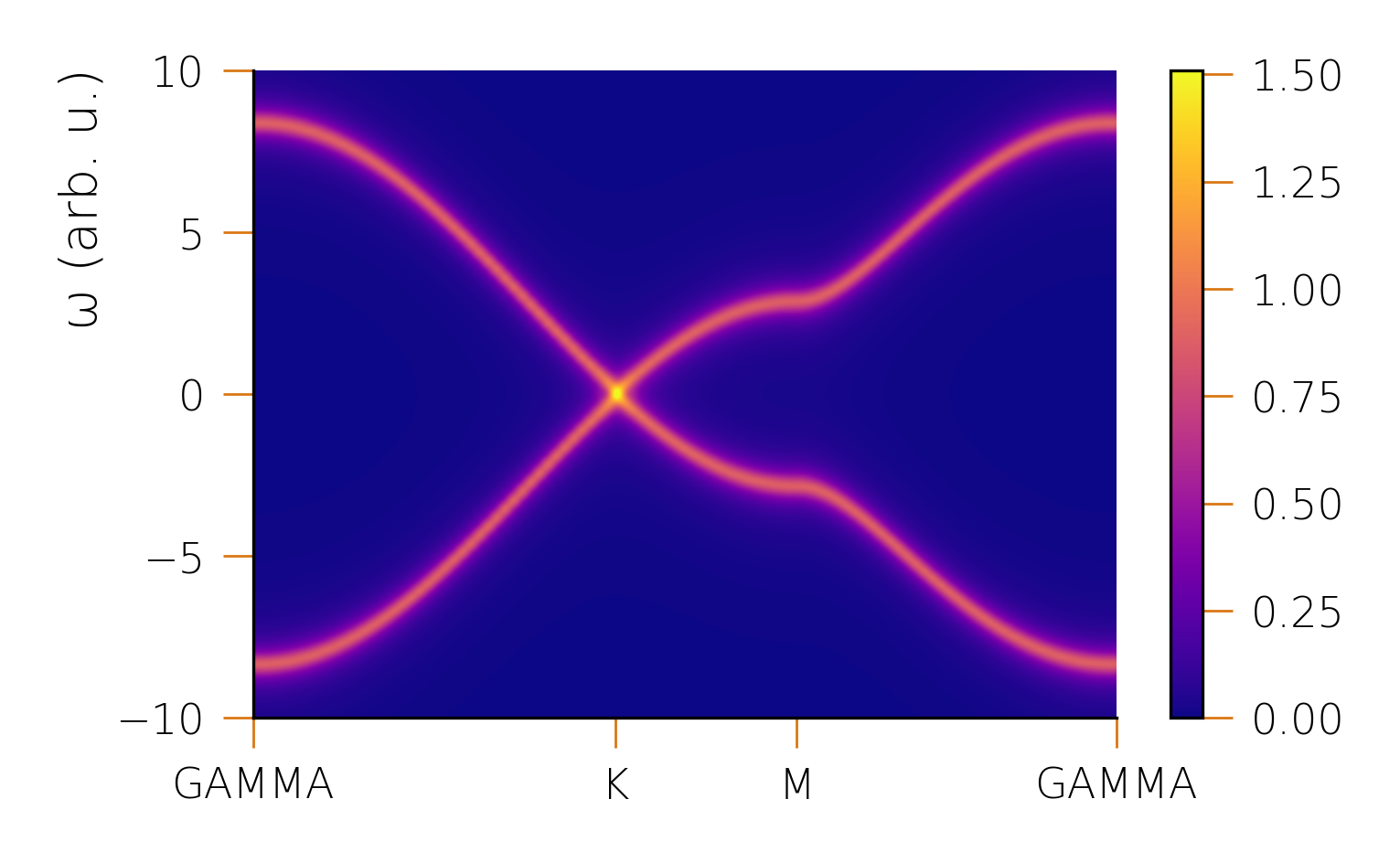}
\caption{
Single-particle spectral function obtained via cluster perturbation theory for an interacting $4\times3$ graphene flake with periodic boundary conditions and interaction strengths $U_1=0.2$, $U_2=0.1$, and $U_3=0.05$. 
The Dirac cone remains clearly visible and no gap opens at the Fermi energy. 
The system remains gapless, confirming that the observed dynamics arises from many-body scattering rather than interaction-induced gap formation.
}
    \label{fig:A3}
\end{figure}

In this appendix we verify that the interacting tight-binding model studied in the main text does not exhibit a gap at the Fermi energy for the chosen interaction parameters. To this end, we compute the single-particle spectral function using cluster perturbation theory \cite{enenkel2023, Senechal2000} with a $4\times3$ graphene flake with hard-wall boundary conditions.

Cluster perturbation theory provides an approximate reconstruction of the momentum-resolved spectral function of the infinite system from the exact solution of a finite interacting cluster. In the present case, the cluster corresponds to the finite graphene flake used in the time-evolution simulations, allowing for a consistent comparison between equilibrium spectral properties and nonequilibrium dynamics.

The resulting band structure is shown in Fig.~\ref{fig:A3}. The characteristic Dirac cone at the Fermi energy is clearly preserved, and no gap opening is observed. While interactions lead to a finite broadening of spectral features and a slight renormalization of the dispersion, the low-energy structure remains gapless.

This observation is important for the interpretation of the quench dynamics discussed in the main text. In particular, it confirms that the relaxation processes occur in a gapless system with a Dirac-like low-energy spectrum, and are therefore not influenced by interaction-induced gap formation. The correlation effects observed in the orbital entropy and excitation-space analysis thus originate from many-body scattering processes rather than from a qualitative change of the underlying single-particle spectrum.

\section{Finite-size effects in excitation-space truncations}
\label{app:finitesize_excitation}

In the main text, the particle--hole excitation spaces $\VV_\ell$ were initialized from the single reference state $\ket{\xi_0}$, cf.~\Eqref{eq:single_state_V0}. 
Here we examine how this choice affects the apparent convergence with excitation order by instead starting from an enlarged initial space,
\begin{equation}
  \VV^*_0 = \{\ket{\Psi_0},\, \ket{\xi_0}\},
  \label{eq:two_state_V0}
\end{equation}
which includes both the non-interacting ground state and the particle--hole excited state. 
The corresponding $n$-particle--hole excitation spaces are then constructed recursively according to \Eqref{eq:Vn_recursion}.

\begin{figure}
    \includegraphics[width=0.33\textwidth]{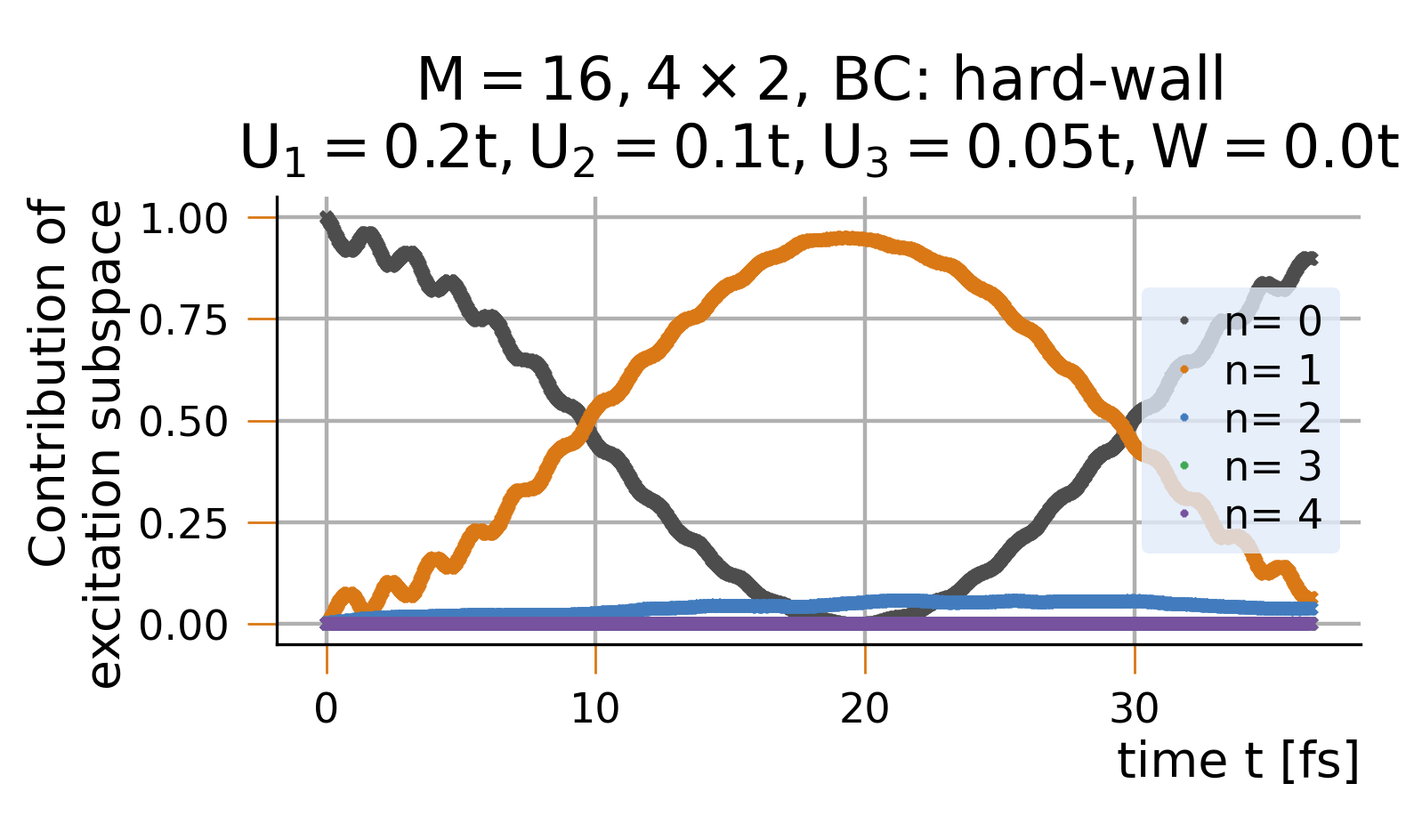}
  \includegraphics[width=0.33\textwidth]{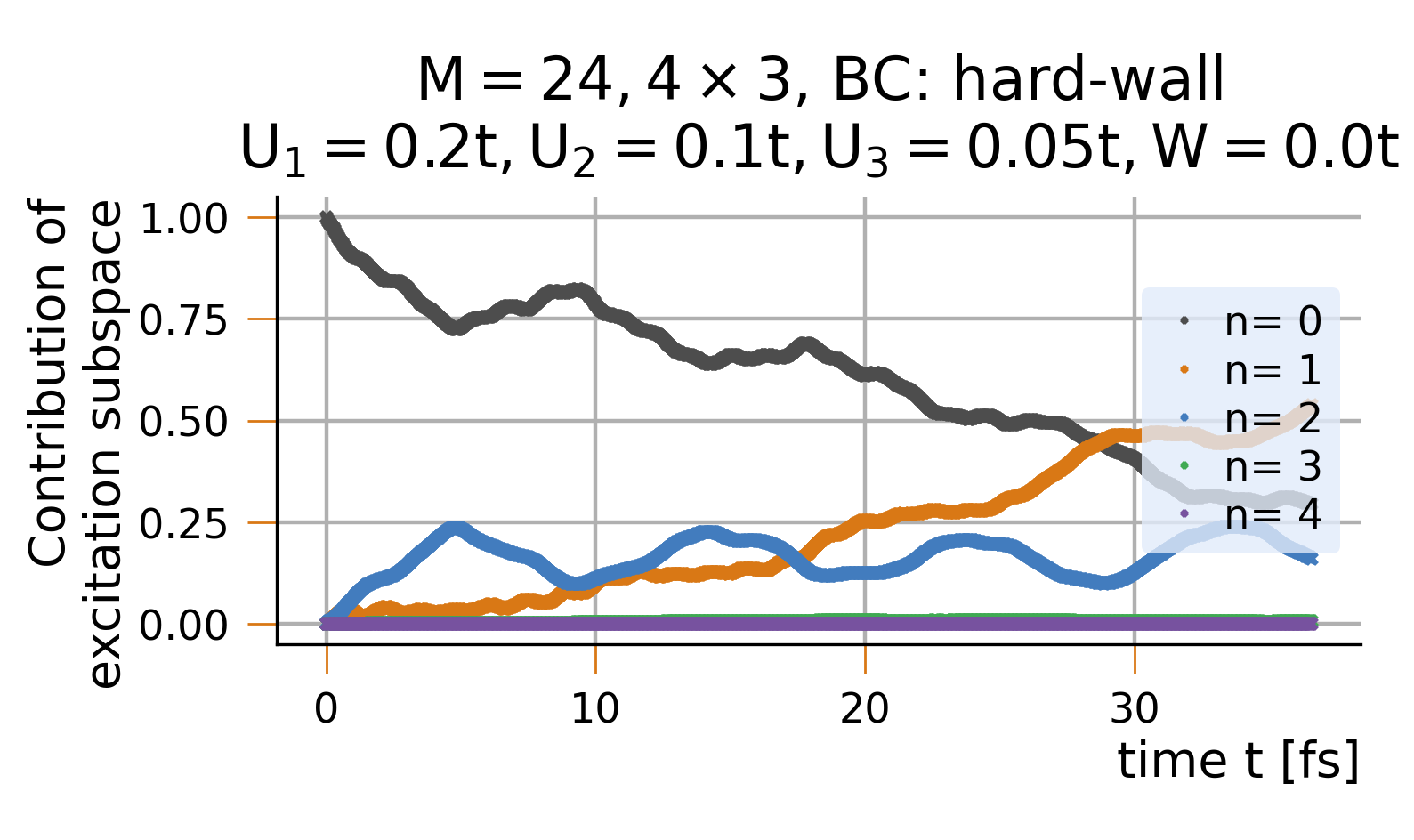}
  \includegraphics[width=0.33\textwidth]{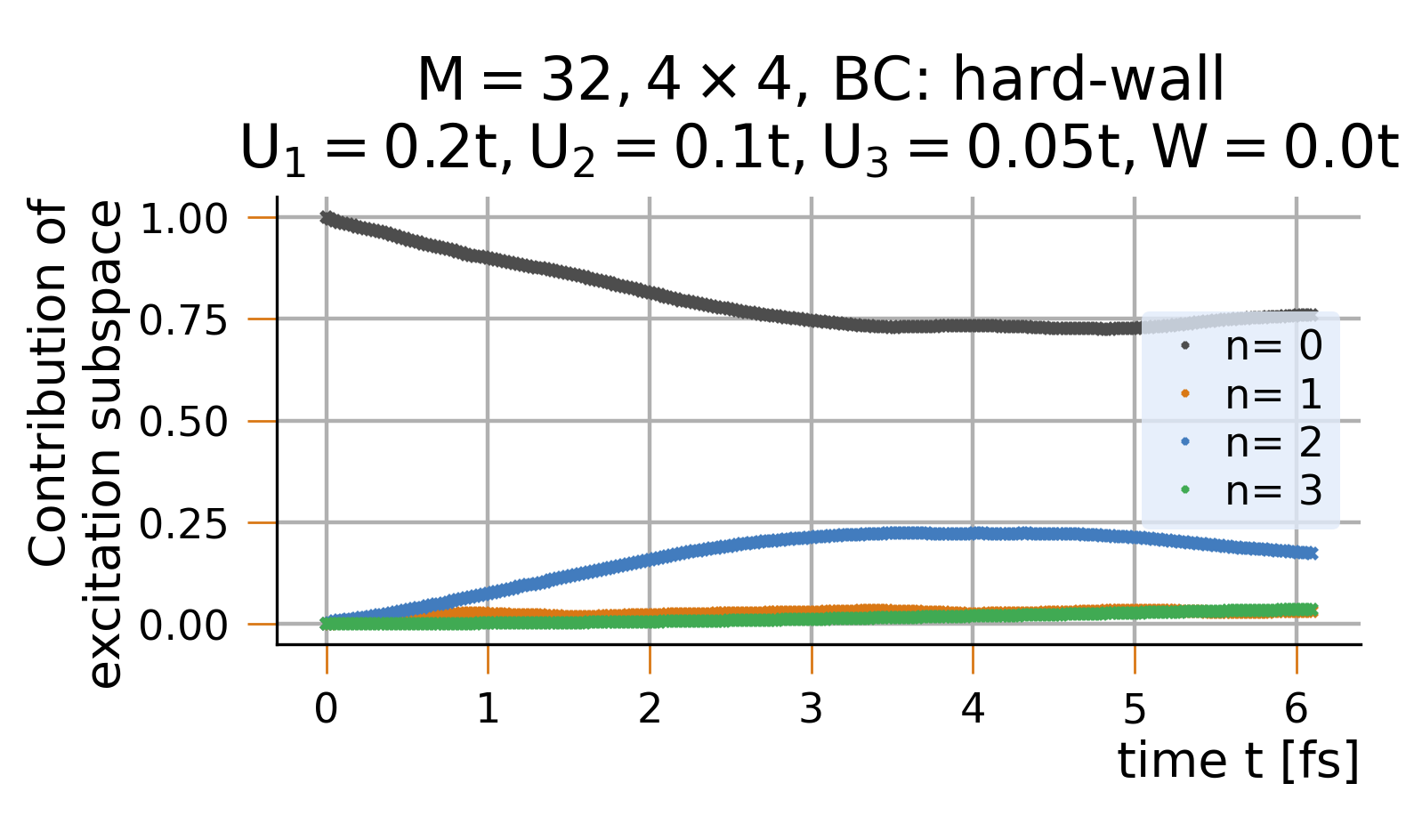}
\caption{
Contribution of excitation sectors for increasing system size at $U_1=0.2$, $U_2=0.1$, and $U_3=0.05$ and without disorder ($W=0$). 
Left: $4\times2$ system (16 sites), middle: $4\times3$ system (24 sites, calculated within $\VV_4$), right: $4\times4$ system (32 sites, calculated within $\VV_3$). 
While the smallest system suggests rapid convergence at low excitation order, larger systems reveal substantial contributions from higher-order sectors.
}
\label{fig:Vinit2}
\end{figure}

Figure~\ref{fig:Vinit2} shows the contributions of different excitation sectors for systems of increasing size. 
For the smallest system ($4\times2$), one might conclude that restricting the dynamics to the single particle--hole excitation subspace already provides an adequate description when starting from $\VV_0^*$. 
However, this conclusion does not hold for larger systems: already for the $4\times3$ lattice, higher-order excitation sectors contribute significantly, demonstrating that the apparent convergence at low excitation order is a finite-size artefact.

\begin{figure}
    \includegraphics[width=0.45\textwidth]{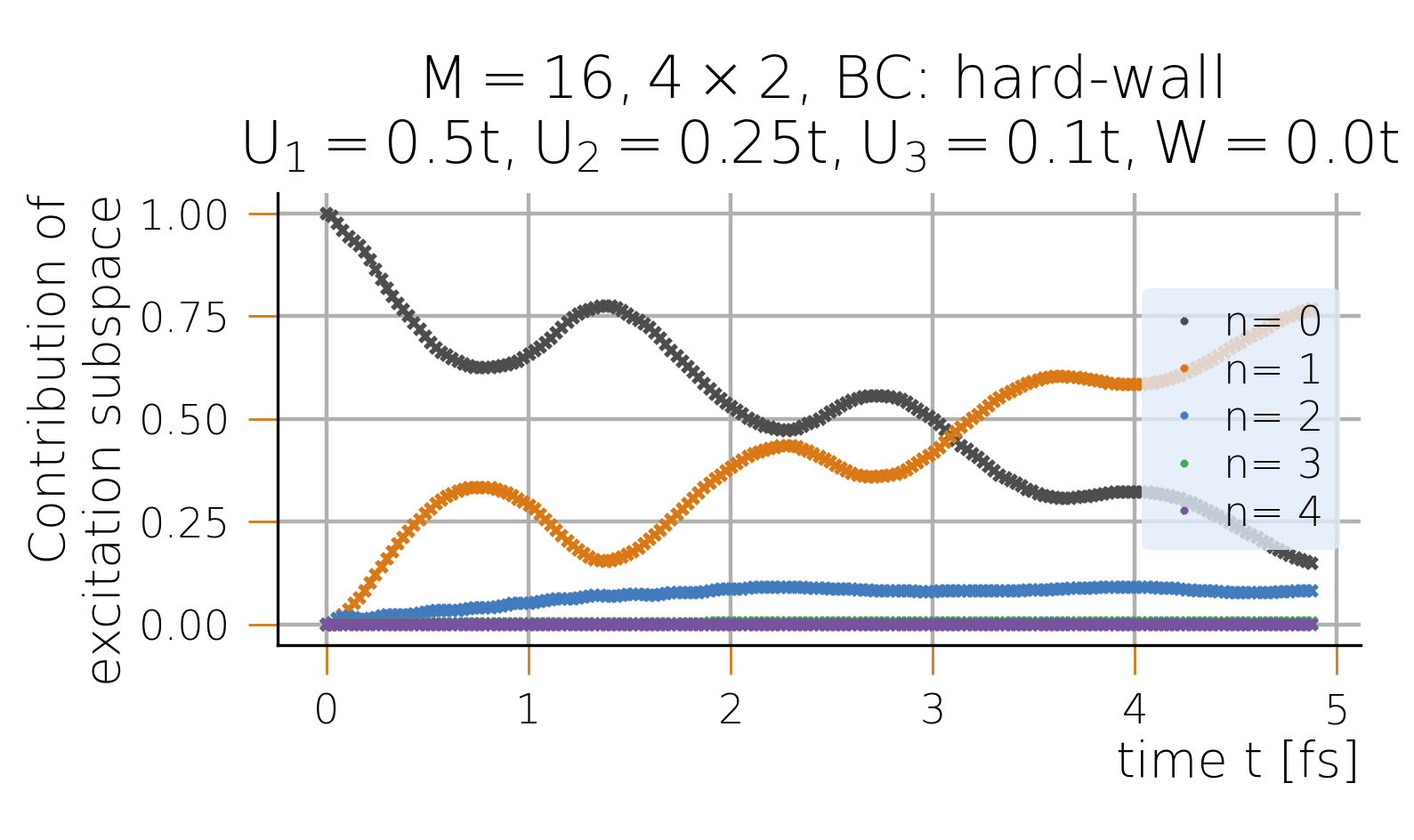}
  \includegraphics[width=0.45\textwidth]{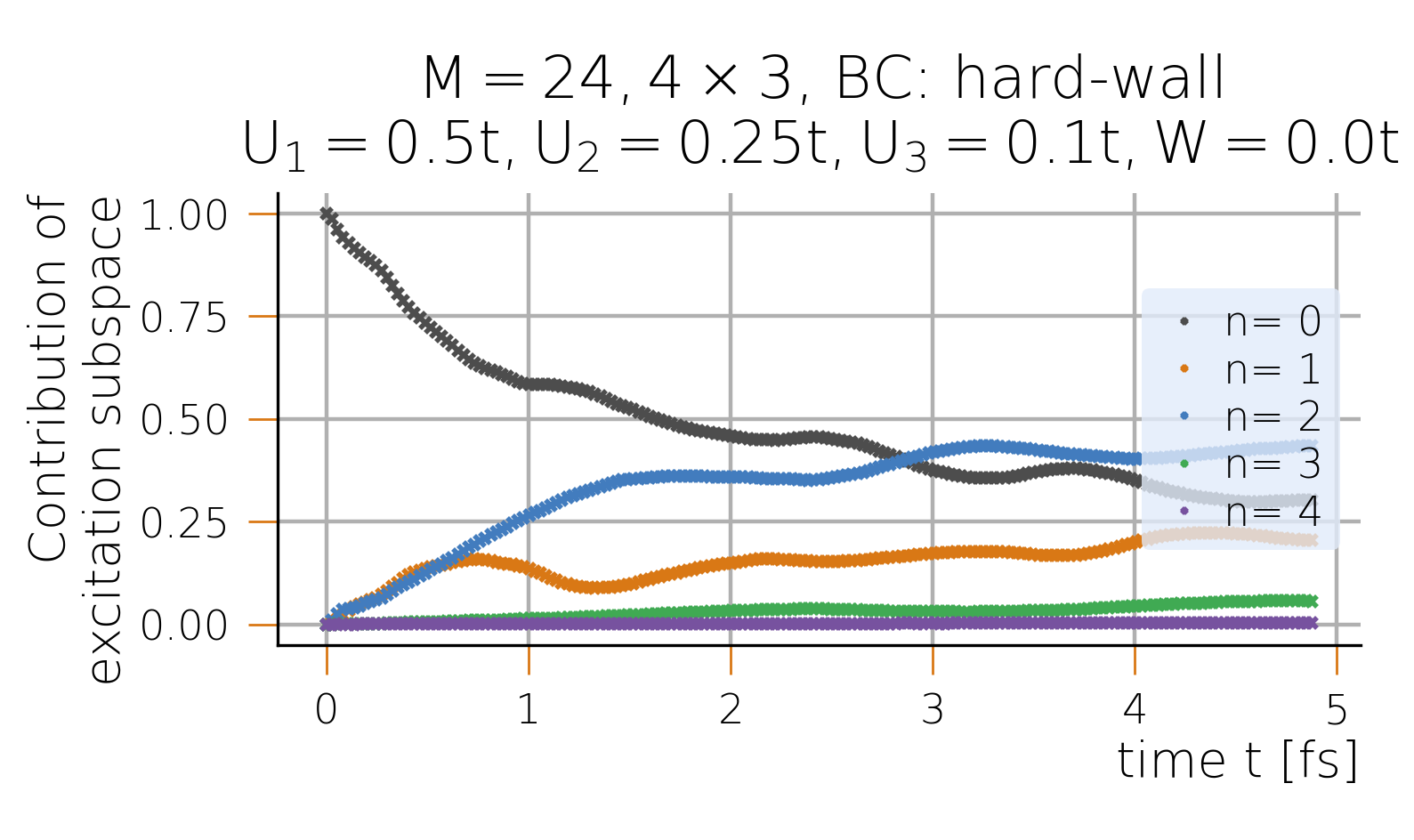}
\caption{
Same analysis as in Fig.~\ref{fig:Vinit2} for stronger interactions ($U_1=0.5$, $U_2=0.25$, $U_3=0.1$) without disorder ($W=0$). 
Finite-size effects become more pronounced, and larger systems show clear contributions from higher-order excitation sectors.
}
\label{fig:Vinit2_largeU}
\end{figure}

This effect becomes even more pronounced at stronger interaction strengths, as shown in Fig.~\ref{fig:Vinit2_largeU}. 
While the smallest system still suggests a dominant role of low-order excitations, larger systems clearly exhibit substantial weight in higher-order sectors. 
This highlights that finite-size effects can strongly bias the perceived importance of excitation order and may lead to misleading conclusions if not carefully controlled.

\end{appendix}





\bibliography{References.bib}


\end{document}